\documentclass[10pt, conference, letterpaper]{IEEEtran}
\IEEEoverridecommandlockouts
\usepackage{cite}
\usepackage{amsmath,amssymb,amsfonts}
\usepackage{algorithmic}
\usepackage{graphicx}
\usepackage{multirow}
\usepackage{textcomp}
\usepackage{xcolor}
\usepackage{url}

\usepackage[caption=false,font=footnotesize,labelfont=rm,textfont=rm]{subfig}
\def\BibTeX{{\rm B\kern-.05em{\sc i\kern-.025em b}\kern-.08em
		T\kern-.1667em\lower.7ex\hbox{E}\kern-.125emX}}

\begin{document}
	
	\title{Towards Robust RF Fingerprint Identification Using Spectral Regrowth and Carrier Frequency Offset}

	\author{
		\IEEEauthorblockN{
			Lingnan Xie\IEEEauthorrefmark{1},
			Linning Peng\IEEEauthorrefmark{1}\IEEEauthorrefmark{2},\thanks{Lining Peng is the corresponding author. This work was supported in part by the National Natural Science Foundation of China under Grant 62171120.}
			and Junqing Zhang\IEEEauthorrefmark{3}
			}

		\IEEEauthorblockA{
			\IEEEauthorrefmark{1}
			School of Cyber Science and Engineering, Southeast University, Nanjing, China
			}
		
		\IEEEauthorblockA{
			\IEEEauthorrefmark{2}
			Purple Mountain Laboratories for Network and Communication Security, Nanjing, China
			}
		\IEEEauthorblockA{
			\IEEEauthorrefmark{3}
			Department of Electrical Engineering and Electronics, University of Liverpool, Liverpool, United Kingdom
			}
        \IEEEauthorblockA{
			Email:\{lnxie,pengln\}@seu.edu.cn, junqing.zhang@liverpool.ac.uk
			}
	}


\maketitle

\begin{abstract}
Radio frequency fingerprint identification (RFFI) is a promising device authentication approach by exploiting the unique hardware impairments as device identifiers. Because the hardware features are extracted from the received waveform, they are twisted with the channel propagation effect. Hence, channel elimination is critical for a robust RFFI system. In this paper, we designed a channel-robust RFFI scheme for IEEE 802.11 devices based on spectral regrowth and proposed a carrier frequency offset (CFO)-assisted collaborative identification mechanism. In particular, the spectral regrowth was utilized as a channel-resilient RFF representation which is rooted in the power amplifier nonlinearity. While CFO is time-varying and cannot be used alone as a reliable feature, we used CFO as an auxiliary feature to adjust the deep learning-based inference. Finally, a collaborative identification was adopted to leverage the diversity in a multi-antenna receiver. Extensive experimental evaluations were performed in practical environments using 10 IEEE 802.11 devices and a universal software radio peripheral (USRP) X310 receiver with 4 antennas. The results demonstrated the effectiveness of the proposed method against diverse channel conditions and CFO drift, where an average classification accuracy of 92.76\% was achieved against channel variations and a 5-month time span, significantly outperforming existing methods.

	
\end{abstract}


\section{Introduction}
\IEEEPARstart{T}{he} Internet of Things (IoT) has become an indispensable technology for modern life, enabling numerous practical applications such as medical monitoring \cite{9961104} and industrial control \cite{9195014}. At the same time, the growing number of IoT devices and the openness of wireless communication underscore the need for reliable authentication methods to safeguard against unauthorized network access by malicious users \cite{9149584}. Traditional authentication methods primarily rely on software addresses such as Internet Protocol and/or Media Access Control (MAC) addresses. However, they may be vulnerable to spoofing attacks~\cite{8737455}.

Radio frequency fingerprint identification (RFFI) has been evolving rapidly recently. This technique utilizes the distinctive hardware imperfections contained in the received RF signal, which are naturally generated during the manufacturing process and therefore difficult to imitate \cite{refId0}. Additionally, since the imperfections of low-cost electronic components tend to be distinguishable, the RFF is particularly suitable for IoT device identification. Therefore, extensive researches have been conducted, covering a diverse range of IoT techniques, including ZigBee \cite{9517121, 10508246,7464336}, LoRa \cite{9715147, 9965952}, Wi-Fi \cite{10050441, 10430094, 9979789, 9335608, 8882379, 9155259, 10332157}, LTE \cite{10360105}, ADS-B \cite{10511278,10285131}, and Bluetooth~\cite{10107741,10229972}.

Deep learning has been widely used in RFFI, which consists of two stages, namely training and inference. During the training stage, a receiver is employed to collect adequate signal packets through wireless channels from the devices to be registered. Then, a preprocessing step will be conducted, followed by the extraction of the RFF representations and the training of the classifier. During the inference stage, the same type of RFF will be extracted from the newly received signals which are also affected by the wireless channel, and the identity will be inferred with the trained classifier.

Notably, the channel impacts will affect the data distribution of RFF in the training and test set, which may lead to overfitting issues during the training stage, thereby deteriorating the recognition result \cite{9155259}, \cite{9348261}. Therefore, channel effect elimination is important.
A typical method is to design tailored signal processing algorithms to eliminate channel effects, which can be achieved by exploring the similarity of channel impacts and the differences in RFFs within the coherence time~\cite{9715147,9979789,8882379} or the coherence bandwidth \cite{10050441}. In \cite{9715147}, the multiplicative channel interference of LoRa signal was mitigated by dividing the adjacent columns of the spectrum obtained by short-time Fourier transform (STFT).
Additionally, for broadband signals such as Wi-Fi which are inherently sensitive to multi-path fading, there is a heightened requirement for channel cancellation to derive a reliable RFF. In \cite{9979789}, the difference between the logarithmic spectrum of two adjacent symbols was used as RFF. Besides, the study in \cite{10050441} proposed the quotient of the spectrum of adjacent subcarriers as a channel-resilient feature. 
In \cite{8882379}, traditional channel estimation and equalization based on the least-square (LS) method were leveraged to mitigate the channel impacts. 
However, since the common parts of the RFF would be removed along with the channel impacts during the above processes, there will be non-negligible feature loss.

While signal processing is useful in mitigating channel effects, it is equally crucial to explore hardware impairments that exhibit resilience against channel variations. In this regard, the power amplifier (PA) nonlinearity is a promising feature. In \cite{10430094} and \cite{5740592}, different estimation algorithms were proposed to extract nonlinearity features based on memory and memoryless PA models, respectively. However, the estimation process may require empirically set parameters. In addition, the spectral regrowth,  exhibited as spurious emissions on the spectrum, is generated by the PA nonlinearity \cite{1703853} and can be seen as an RFF representation of the nonlinearity feature.
Therefore, it is promising to use it for device identification. The spectral regrowth of WLAN transmitter was measured in \cite{7122344} as one of the features that are independent of data symbols. However, the effectiveness of spectral regrowth as an RFF representation was not fully analyzed, and there was a lack of validation in real-world scenarios. In~\cite{10229972}, the spectral regrowth controlled by a physically unclonable function (PUF) was utilized for device identification, yet the effect of channel variation was not taken into consideration.

Besides PA nonlinearity, CFO has also been used as a channel-independent feature, which was directly involved in the device identification process \cite{6804410,8485917}. However, due to the proven time-varying nature of CFO in IoT devices~\cite{10184130}, using CFO alone tends to be unreliable. Therefore, the utilization of CFO has shifted towards serving as an auxiliary feature to aid the recognition based on the relative stability over a short period of time and the coarse distinguishability across devices~\cite{9448147,8882379}. However, in the case of Wi-Fi devices, there is no research on the impact of CFO drift on identification results over long time spans (e.g., several months), let alone providing an algorithm that resists drift. 





In this paper, a channel-robust RFFI framework is proposed, and Wi-Fi is chosen as a case study. Specifically, we explore the potential of spectral regrowth as an RFF representation to extract channel-resilient features, and a collaborative classification method is proposed based on a multi-antenna receiving system. Moreover, a CFO-assisted method is proposed to resist CFO drift and aid the identification. Extensive experiments were conducted in various environments. The main contributions of this work are summarized as follows.
\begin{itemize}
	\item We modelled the spectral regrowth based on the memory effect and nonlinearity of PA characteristics, together with the differences between the RFF representations based on spectral regrowth and the spectrum on active subcarriers of Wi-Fi signals. A 128-length RFF feature with one dimension was constructed, and an extraction method was also presented.

	\item We designed a CFO-assisted collaborative classification method by exploiting the softmax scores of the multiple received samples from a single frame of transmitted signal, and utilizing the estimated CFO values to calibrate the prediction results. Specifically, we employed a 4-channel USRP receiver to capture the spatial diversity across individual frames. Notably, the widespread multi-antenna setups in access points (APs) indicate that the proposed algorithm can be seamlessly integrated into existing AP systems.

	\item We carried out extensive experiments under various environments and a collection time span of several months. In detail, static scenarios with line-of-sight (LOS) and none-line-of-sight (NLOS) configurations, as well as the dynamic scenario with moving transmitters were involved. 
    In order to promote research in this field, the collected experimental dataset of Wi-Fi devices is openly available online\footnote{\url{https://ieee-dataport.org/documents/wi-fi-dataset-channel-robust-rffi}}.
    \item We experimentally compared the effectiveness of four RFF representations against channel variations, namely the spectral regrowth on inactive subcarriers, the spectrum on active subcarriers, and two related works. It was found out that the spectral regrowth could reach an average accuracy of 92.76\%, which was far superior to the others with an improvement of at least 8.94\%. Furthermore, the experimental results proved the effectiveness of the proposed CFO-assisted collaborative classification model, where the average accuracy was improved from 68.61\% to 92.76\% for the spectral regrowth-based scheme. Besides, when faced with CFO drift caused by a long time span, the proposed CFO-assisted method was still able to improve the recognition accuracy by 9.25\% while the related methods failed to achieve positive effects.
\end{itemize}

The remainder of this paper is organized as follows. Section \ref{sec:preliminary} briefly introduces the preamble structure and the spectrum composition of IEEE 802.11 signals. Section \ref{sec:sys_overview} provides an overview of the proposed RFFI system. Section \ref{sec:representation}  demonstrates the differences between the RFF representations on active and inactive subcarriers, and presents the extraction method. Section \ref{sec:collaborative classification} proposes the CFO assisted collaborative classification mechanism. Experimental evaluation results of the system performance are demonstrated in Section \ref{sec:experiments}, and finally Section \ref{sec:conclusion} concludes this paper.

\section{Preliminary: IEEE 802.11 Signal}\label{sec:preliminary}

\begin{figure}[!t]
	\centering
	\includegraphics{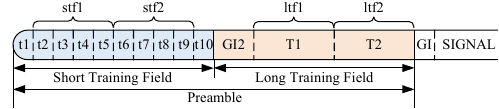}
	\caption{Preamble structure of IEEE 802.11 signal.}
	\label{fig:preamble structure}
\end{figure}

The IEEE 802.11 standard defines a preamble adopting orthogonal frequency division multiplexing (OFDM) modulation at the beginning of every packet. As shown in Fig.~\ref{fig:preamble structure}, the time-domain preamble consists of a short training field (STF) and a long training field (LTF), where the STF comprises ten repetitive short training symbols (STSs), while the LTF is composed of one cyclic prefix (CP) and two long training symbols (LTSs). The STSs, LTSs, and CP are denoted as \{t1, t2,..., t10\}, \{T1, T2\}, and GI2, respectively.

Furthermore, for frequency domain analysis, since the spectrum calculation of the first and the last STS (i.e., t1 and t10) may be interfered with by the neighboring segments of signal in the time domain, we concatenate the t2-5, t6-9 from the STF as two OFDM symbols. Moreover, we choose the T1, T2 from the LTF as the other two available OFDM symbols in the preamble. For convenience, these four symbols are denoted as \{stf1, stf2, ltf1, ltf2\}.

In the frequency domain, there are 64 subcarriers over a 20-MHz OFDM channel. The STS only uses 12 of them with indices
\begin{equation}
	\zeta_{STF}\!=\![-24,\!-20,\!-16,\!-12,\!-8,\!-4,4,8,12,16,20,24],
	\label{eq:ISTF}
\end{equation}
while the LTS utilizes 52 active subcarries, yielding
\begin{equation}
	\zeta_{LTF}=\zeta_{active}=[-26,-1]\cup[1,26],\zeta_{active}\in Z.
	\label{eq:Iactive}
\end{equation}

\section{System Overview}\label{sec:sys_overview}
\begin{figure}[!t]
	\centering
	\includegraphics[width=3.5in]{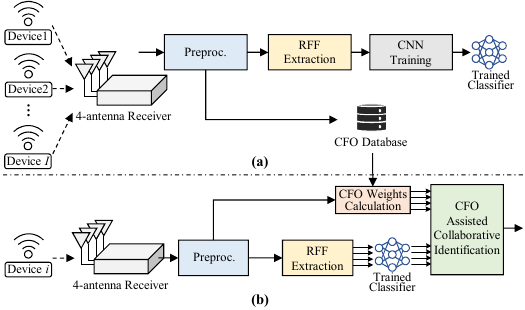}
	\caption{The proposed RFFI system. (a) Training stage. (b) Inference stage.}
	\label{fig:flowchart}
\end{figure}

The proposed RFFI system is depicted in Fig.~\ref{fig:flowchart}, which utilizes the preamble of the received Wi-Fi signal and a deep learning-based classifier to accomplish the device classification. Specifically, the system involves a training stage and an inference stage. In the training stage, the collected signal packets are labelled and preprocessed, and then the RFF representation is extracted from the preamble parts of them, which is subsequently fed into the deep learning model to accomplish the neural network training. Upon the completion of the training stage, the system turns into the inference stage, at which point it takes on the function of recognizing newly received Wi-Fi signals.

\subsection{Received Signal}
As illustrated in Fig.~\ref{fig:flowchart}, considering a multi-output wireless communication scenario with $I$ transmitters and a receiver with $K$ antennas, the time-domain signal from the $i$-th transmitter received by the $k$-th antenna can be described as
\begin{equation}
	y_{i,k}(t)=h(t)*\mathcal{G}_i\bigl(x(t)e^{j2\pi f_c^{i} t}\bigr)+n(t),
	\label{eq:yt}
\end{equation}
where $h(t)$ is the channel effect, $*$ represents linear convolution operation, $\mathcal{G}_i(\cdot)$ indicates the distortions of the power amplifier, $x(t)$ is the input ideal signal, $f_c^{i}$ represents the carrier frequency of the $i$-th transmitter, and $n(t)$ is the additive white Gaussian noise. In addition, the number of available antennas is $K=$ 4 in this paper.

While there are various hardware impairments such as I/Q imbalance, oscillator imperfection, etc~\cite{9450821}. In this paper, we focus on PA characteristics and the frequency offset of Wi-Fi devices.

\subsection{Signal Preprocessing}
As shown in Fig.~\ref{fig:flowchart}, a preprocessing step is set up in both the training and inference stages. Specifically, for the received signal, a standard synchronization algorithm as in \cite{8031977} is firstly employed to precisely find the start point of the packet. Then, the preamble part is extracted, followed by the frequency offset compensation \cite{8798719} which aims to prevent it from affecting the RFF extraction. Notably, the estimated CFO values are stored in the database, which will be utilized later in the identification process. Finally, the power normalization is carried out.

\subsection{RFF Extraction}
After preprocessing, the preamble of the signal is transformed into a frequency domain expression, yielding
\begin{equation}
	Y_{i,k}(\omega)=H(\omega)S_{i}(\omega)+N(\omega), 
 \label{eq:Yiomega}
\end{equation}
where $Y_{i,k}(\omega)$, $H(\omega)$, $S_{i}(\omega)$, and $N(\omega)$ denote the frequency domain description of the preprocessed received signal, the channel impact, the output of the power amplifier, and the additive noise, respectively. 

Evidently, the RFFs are embedded in $Y_{i,k}(\omega)$, thus the primary objective is to find an appropriate RFF representation within the frequency domain that can facilitate the classifier's capacity to learn the unique hardware features. Details can be found in Section~\ref{sec:simulation} and Section~\ref{sec:signal_representation}.

\subsection{CFO Assisted Collaborative Identification}\label{sec:overview_collaborative}
During the training stage, as illustrated in Fig.~\ref{fig:flowchart}(a), a convolutional neural network (CNN) model is employed as the classifier, which is carefully designed to mine out latent features from the RFF representations that are extracted from the signal samples collected in a dynamic scenario with moving transmitters. Notably, all the samples received from the four antennas are fed into the classifier during training.

In the inference stage, as shown in Fig.~\ref{fig:flowchart}(b), taking advantage of the 4-antenna receiver, a decision fusion method is employed based on different predictions derived by the trained classifier for the multiple received signal samples.

Moreover, the estimated CFO values are incorporated into the identification process, which has been proven to be effective for device identification when using the appropriate algorithms despite the time-variant nature of CFO. Details of the identification algorithm can be found in Section~\ref{sec:collaborative classification}.

\section{RFF Extraction}\label{sec:representation}
\subsection{Nonlinearity and Memory Effect of Power Amplifier}\label{sec:nonlinear}
A nonlinear system with $M$-tap memory is generally represented by the Volterra series, which can be described as a sum of multidimensional convolutions in discrete time as follows:
\begin{equation}
	\mathcal{G}_i\bigl(u(t)\bigr)=\sum_{d=1}^{D}g_d(t), \label{eq:Gut}
\end{equation}
where
\begin{equation}
	g_d(t)=\sum_{m_1=0}^{M-1}\cdots\sum_{m_d=0}^{M-1}\alpha_d(m_1,\cdots,m_d)\prod_{l=1}^{d}u(t-m_l). \label{eq:gdt}
\end{equation} 
Here, $u(t)$ is the input $x(t)e^{j2\pi f_c^{i} t}$, while $D$ and $M$ are the largest dimension and memory length, respectively, and $\alpha_d(m_1,\cdots,m_d)$ denotes the $d$th-order Volterra kernel.

Specially, for $d=1$, there is
\begin{equation}
	g_1(t)=\sum_{m_1=0}^{M-1}\alpha_1(m_1)u(t-m_1)=\alpha_1(t)*u(t),
\end{equation}
which can be described in the frequency domain as
\begin{equation}
	S_1(\omega)=A_1(\omega)U(\omega), \label{eq:S1omega}
\end{equation}
where $A_1(\omega)$ and $U(\omega)$ denote the discrete-time Fourier transform (DTFT) of $\alpha_1(t)$ and $u(t)$, respectively.

According to (\ref{eq:S1omega}), it is noteworthy that when $d=1$, the system only exhibits a linear influence on the input frequency-domain signal, which is dominated by the memory effect, i.e., multiplying the input $U(\omega)$ by the coefficient $A_1(\omega)$, whereas the nonlinear features have not yet contributed to the output.

When $d\ge2$, the change of coordinates has been proven useful to perform frequency domain analysis of nonlinear systems modeled by the Volterra series \cite{651189}, written as
\begin{equation}
	m_l=\begin{cases} s, & l=1 \\s+r_{l-1}, & l=2,\cdots,d \end{cases}
\end{equation}
The output of the $d$th-order Volterra kernel in (\ref{eq:gdt}) can then be expressed as a sum of one-dimensional (1-D) convolutions, which may be written as
\begin{align}	
g_d(t)=&\sum_{r_1=0}^{M-1}\cdots\hspace{-0.3cm}\sum_{r_{d-1}=r_{d-2}}^{M-1}\hspace{-0.3cm}\sum_{s=0}^{M-1-r_{d-1}} \hspace{-0.3cm} \beta_{r_1,\cdots,r_{d-1}}(s)v_{r_1,\cdots,r_{d-1}}(t-s)\nonumber\\
=& \sum_{r_1=0}^{M-1}\cdots \hspace{-0.3cm}\sum_{r_{d-1}=r_{d-2}}^{M-1} v_{r_1,\cdots,r_{d-1}}\!(t)\!*\!\beta_{r_1,\cdots,r_{d-1}}\!(t).
			\label{eq:gdt2}
\end{align}
Note that $v_{r_1,\cdots,r_{d-1}}(t)=u(t)\prod_{l=1}^{d-1}u(t-r_l)$, and the 1-D filters $\beta_{r_1,\cdots,r_{d\!-\!1}}\!(t)\!=\!C(0,r_1,\! \cdots \!,r_{d-1})\alpha_d(t, t \! + \! r_1,\! \cdots \!, t \! + \! r_{d-1})$, where $C(0,r_1,\cdots,r_{d-1})=C(s+0,s+r_1,\cdots,s+r_{d-1})=C(m_1,\cdots,m_d)$, which denotes the number of different possible permutations of the set of numbers $m_1,\cdots,m_d$.

Taking the DTFT of (\ref{eq:gdt2}), the frequency domain description of the $d$th-order Volterra kernel output can be modeled as
\begin{align}
	S_d(\omega) \! = \! \sum_{r_1=0}^{M-1}\cdots \! \sum_{r_{d\!-\!1}\!=r_{d\!-\!2}}^{M-1} \!  V_{r_1,\cdots,r_{d\!-\!1}}\!(\omega)B_{r_1,\cdots,r_{d\!-\!1}}\!(\omega),
	\label{eq:Sdomega}
\end{align}
where $V_{r_1,\cdots,r_{d-1}}(\omega)$ and $B_{r_1,\cdots,r_{d-1}}(\omega)$ are the DTFT results of $v_{r_1,\cdots,r_{d-1}}(t)$ and $\beta_{r_1,\cdots,r_{d-1}}(t)$, respectively.

Hence, from (\ref{eq:Gut}), (\ref{eq:S1omega}) and (\ref{eq:Sdomega}), the frequency-domain output signal of the power amplifier can be represented as
\begin{equation}
		\begin{aligned}
			S(\omega)\!=&\sum_{d=1}^DS_d(\omega)\\
			=&A_1(\omega)U(\omega)\\
			+&\!\sum_{d=2}^D\sum_{r_1=0}^{M-1}\!\cdots\!\sum_{r_{d\!-\!1}\!=r_{d\!-\!2}}^{M-1}\!\! V_{r_1,\cdots,r_{d\!-\!1}}(\omega)\!B_{r_1,\cdots,r_{d\!-\!1}}(\omega),
			\label{eq:Somega}
		\end{aligned}	
\end{equation}
which can be regarded as the sum of two elements derived from the system input: a linear transformation result dominated by the memory effect, and a complicated result of the combined influence of both the memory effect and the nonlinearity.

\subsection{RFF Representations for IEEE 802.11 Signal on Active and Inactive Subcarriers}\label{sec:simulation}
For IEEE 802.11 signal, the value of $U(\omega)$ would be zero when $\omega$ belongs to inactive subcarriers, thus in this case the first component in (\ref{eq:Somega}) could be omitted. In contrast, the frequency domain description on active subcarriers is dominated by the first term of (\ref{eq:Somega}). Therefore, although both the active and inactive subcarriers of IEEE 802.11 signal contain fingerprint information, their RFF representations in the frequency domain are different. Specifically, the memory effect dominates the RFFs on active subcarriers, while the nonlinearity accounts for a larger proportion of the spectral regrowth on the inactive subcarriers.

\begin{figure}[!t]
	\centering
	\subfloat[]{\includegraphics[width=3.5in]{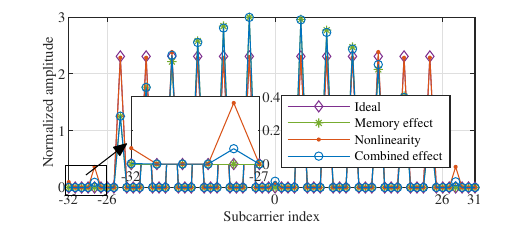}
		\label{fig:nonlinearity_STF}}\\
	\subfloat[]{\includegraphics[width=3.5in]{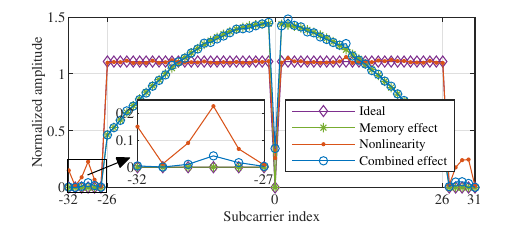}
		\label{fig:nonlinearity_LTF}}
	\caption{Spectrum of the symbol from (a) STF and (b) LTF with/without the effect of the simulated hardware characteristics of the power amplifier.}
	\label{fig:nonlinearity}
	
\end{figure}

In order to better visualize the differences between RFF representations on active and inactive subcarriers, as well as the memory effect and nonlinearity on the signal, the spectrum affected by the hardware characteristics of the amplifier is simulated based on the model presented in Section~\ref{sec:nonlinear}.
Firstly, the identical time-domain IEEE 802.11 OFDM preamble with a channel bandwidth of 20~MHz is generated by the MATLAB Wireless Waveform Generator\footnote{Available: \url{https://www.mathworks.com/help/releases/R2022b/comm/ref/wirelesswaveformgenerator-app.html}}. After that, as shown in Table~\ref{tab:simulation}, Volterra kernels are set up with a largest nonlinearity dimension of 3 and a largest memory length of 2, and the time-domain outputs of the simulated amplifier are calculated as in (\ref{eq:Gut}). Finally, the FFT results of different parameter setup are illustrated in Fig.~\ref{fig:nonlinearity}.

\begin{table}[!t]
	\centering
	\caption{Simulation Setup}
	\begin{tabular}{lccc}
		\hline
		& \multicolumn{1}{c}{Memory effect} & \multicolumn{1}{c}{Nonlinearity} & \multicolumn{1}{c}{Combined effect} \\
		\hline
		$D$ & 1     & 3     & 3 \\
		$M$ & 2     & 1     & 2 \\
		\hline
	\end{tabular}
	\label{tab:simulation}
\end{table}

It can be seen that the memory effect of the power amplifier mainly affects the amplitude on the active subcarriers (i.e., non-zero subcarriers in $[-26,-1]\cup[1,26]$ of STS and LTS), while the nonlinearity results in both the distortions on the active subcarriers and the spectral regrowth on the inactive subcarriers (e.g., $[-32,-27]$) where the ideal amplitude is zero. Moreover, for the spectrum on the active subcarriers, the combined impact of nonlinearity and memory effect results in minor changes in the amplitude compared to the case with only memory effect, which suggests that the memory effect of $A_1(\omega)$ is absolutely dominant on the active subcarriers.


\subsection{RFF Extraction Method}\label{sec:signal_representation}
From the above two subsections, it can be seen that although the amplitude of the spectral regrowth is almost zero, it is promising to utilize it as an RFF representation since its strong correlation to the nonlinearity feature of the power amplifier, and the extraction method is presented as follows.

After obtaining the preprocessed preamble, four symbols are acquired in the time domain as described in Section~\ref{sec:preliminary}, namely \{stf1, stf2, ltf1, ltf2\}, and the frequency domain description of them can be written as (\ref{eq:Yiomega}), where the model for $S_{i}(\omega)$ is presented in Section~\ref{sec:nonlinear}.

Moreover, with the total subcarrier indices $\zeta_{total}=[-32,31],\zeta_{total}\in Z$, and the active subcarriers indices in (\ref{eq:ISTF}) and (\ref{eq:Iactive}), the indices of inactive subcarriers can be written as
\begin{equation}
	\begin{aligned}
		\omega_{STF}=\zeta_{total} \setminus \zeta_{STF}\\
		\omega_{LTF}=\zeta_{total} \setminus \zeta_{LTF}
		\label{eq:omegaSTFLTF}
	\end{aligned}
\end{equation}
where $\setminus$ denotes the set difference operation.

Therefore, combined with the frequency domain representation in (\ref{eq:Somega}), the spectral regrowth generated by the power amplifier on the inactive subcarriers of STF and LTF can be denoted as $S(\omega_{STF})$ and $S(\omega_{LTF})$, respectively.

Finally, a concatenation operation is implemented for turning four bands of spectrum data into a 128-length RFF feature representation with one dimension, where the spectral regrowth of stf1, stf2, ltf1, and ltf2 are sequentially connected.

\section{CFO Assisted Collaborative Identification}\label{sec:collaborative classification}
As illustrated in Fig.~\ref{fig:flowchart}(b), the proposed RFFI system leverages a collaborative classifier based on a CNN model and a decision fusion mechanism. Moreover, the estimated CFO values are utilized to calibrate the prediction results. 

\subsection{Convolutional Neural Network}
\begin{figure}[!t]
	\centering
	\includegraphics[width=3.5in]{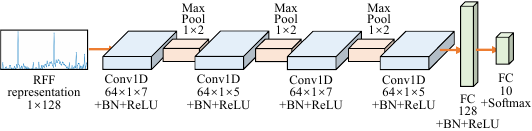}
	\caption{CNN architecture. }
	\label{fig:collaborative classifier}
\end{figure}

After obtaining the RFF representation, a CNN model is employed for the identification process, whose architecture is presented in Fig.~\ref{fig:collaborative classifier}. It consists of four 1-D convolutional layers, the first three of which are followed by a batch normalization (BN) layer and a maxpooling layer, respectively, and the last one of them is connected to two fully connected layers. ReLu function is adopted to activate every convolutional layers and the first fully connected layer, while the softmax function is utilized at the end of the structure.

\subsection{CFO Assisted Decision Fusion}\label{sec:decision fusion}
In this paper, we assume that the MAC addresses of the received signals under test may be forged, and the antenna number $K=4$, so that for a particular frame of transmitted signal, the device identity can only be judged by 4 samples from the 4-antenna receiver. Meanwhile, due to the different channel distortions to which these received samples are subjected, the extracted RFF representations will also have different perturbations, leading to different identification results. However, it is predictable that combining the recognition results of the 4 samples using decision fusion can effectively improve the identification accuracy.

Furthermore, as we will demonstrate in Section~\ref{sec:performance_CFO} and Fig.~\ref{fig:CFOdrift}, although the CFO is time-variant, it remains instructive for RFFI since its relatively stable nature in a short period of time and the coarse distinguishability across devices. Therefore, using the estimated CFOs to calibrate the decision results may be helpful.

Based on the above analysis, a hybrid prediction based on softmax scores and the estimated CFO values is employed.

Given a frame of the transmitted signal from the device under test (DUT) and a receiver with 4 antennas, the prediction results of its 4 received samples, namely $\{y_{1},y_{2},y_{3},y_{4}\}$, can be denoted as
\begin{equation}
	\Psi_{DUT}=\{ \psi_{1},\psi_{2},\psi_{3},\psi_{4} \},
\end{equation}
where $\psi_k$ is the softmax score set of the input RFF, which has $I$ elements reflecting the probability that the RFF belongs to each transmitter.

Besides, as shown in Fig.~\ref{fig:flowchart}(a), a CFO dataset is created for all the $I$ devices in the training stage, and the mean value $\Delta \bar{f}_{i,k}$ of the $i$-th device's CFOs from the $k$-th antenna is calculated. Thus, for the estimated CFO $\Delta f_{DUT,k}$ at the same antenna from the DUT, a weight array can be obtained by
\begin{equation}
	weight_{k}=\begin{bmatrix}
		\frac{1}{\left | \Delta f_{DUT,k}-\Delta \bar{f}_{1,k}\right |},\cdots,\frac{1}{\left | \Delta f_{DUT,k}-\Delta \bar{f}_{I,k}\right |}\end{bmatrix},
    \label{eq:weight}
\end{equation}
which can calibrate the prediction result when $\Delta f_{DUT,k}$ and $\Delta \bar{f}_{i,k}$ are far away from each other. 

Then the softmax score set $\psi_{k}$ is modified to
\begin{equation}
	\psi'_{k}=weight_{k} \cdot \psi_{k}.
	\label{eq:psi'}
\end{equation}
A merged prediction score set $\hat{\psi}_{DUT}$ can be acquired by averaging the $4$ modified score sets, given as
\begin{equation}
	\hat{\psi}_{DUT}=\frac{1}{4}\sum_{k=1}^{4} \psi'_{k}.
\end{equation}

Then the predicted label of DUT can be derived by finding the index with the highest score, formulated as
\begin{equation}
	label={{\arg\max}}(\hat{\psi}_{DUT}).
\end{equation}


\section{Experimental Evaluation}\label{sec:experiments}

\subsection{Experimental Settings}
\subsubsection{Experimental Platform Setup}
\begin{figure}[!t]
	\centering
	\subfloat[]{\includegraphics[width=1.7in]{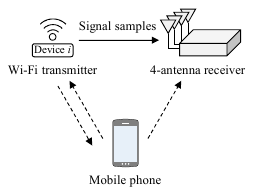}}
	\subfloat[]{\includegraphics[width=1.7in]{./figures/data_collection_platform}}
	\caption{(a) Signal collection setup. (b) Photo of the data collection platform.}
	\label{fig:experimental_settings}
\end{figure}

We employed 10 commercial Wi-Fi devices and a universal software radio peripheral (USRP) X310 with 2 TwinRx RF Daughterboards as the receiver. A mobile phone was used to trigger the signal transmission. The models of the Wi-Fi devices are given in Table~\ref{tab:modelofAPs}. The carrier frequency and bandwidth of DUTs, and the sampling rate of the receiver were configured as 5.825~GHz, 20~MHz, and 20~MS/s, respectively. Additionally, four antennas were arranged in a square configuration with a side length larger than the half-wavelength of the transmitted signal, thus ensuring that the channel conditions among the four antennas were different. The layout of the signal collection scenario and the photograph of the data collection platform are presented in Figs.~\ref{fig:experimental_settings}(a) and (b), respectively.
\begin{table*}[!t]
	\caption{Wi-Fi Devices Used in Experiments\label{tab:modelofAPs}}
	\centering  
	\begin{tabular}{|c|c|c|c|c|c|c|c|c|c|c|}  
		\hline  
		Label & D1 & D2 & D3 & D4 & D5 & D6 & D7 & D8 & D9 & D10 \\ \hline
		Manufacturer & 360 & China Mobile & MERCURY & Motorola & \multicolumn{3}{c|}{TP-LINK} & Tenda & \multicolumn{2}{c|}{LEGUANG} \\ \hline 
		Model & 360P4C & HN140 & MOAP1200D & MWR03 & TL-WDR4310 & \multicolumn{2}{c|}{TL-AP1201P} & AC6 & \multicolumn{2}{c|}{N595} \\ \hline 
	\end{tabular}
\end{table*}

\begin{figure}[!t]
	\centering
	\subfloat[]{\includegraphics[height=1.1056in]{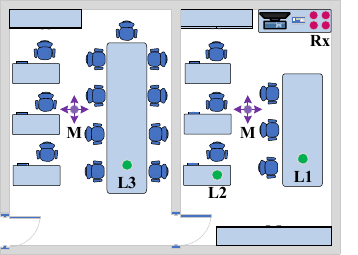}}
	\subfloat[]{\includegraphics[height=1.1056in]{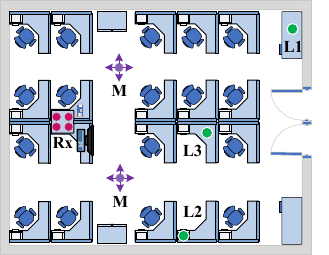}}
	\subfloat[]{\includegraphics[height=1.1056in]{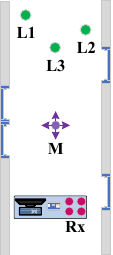}}
	\caption{Layout of different environments. (a) E1. (b) E2. (c) E3.}
	\label{fig:environment_settings}
\end{figure}

The signal preprocessing and RFF extraction were carried out on a PC with a CPU of 13th Gen Intel Core i9-13900K, while the classification was implemented on the same PC with a GPU of NVIDIA Geforce RTX 3090. MATLAB R2022b and Pytorch were utilized.

\subsubsection{Experimental Environments}
Three environments that are consistent with the practical application scenarios of Wi-Fi devices were selected for signal collection. Besides, the time span across different environments was 5 months.

Environment 1 (E1) is an office with a wall in the center, and there are several tables, chairs, cabinets, and computers which may make the multipath effect severe. E2 is another office without a central wall, yet equipped with lots of tables with partitions. Moreover, E3 is a narrow corridor with no furniture in it. The layout of these three environments are illustrated in Figs.~\ref{fig:environment_settings}(a)-(c). Notably, during data collection, the 4-antenna receiver was fixed at location Rx corresponding to the environment configuration.

\subsubsection{Experimental Scenarios}
Based on whether the transmitters were fixed or mobile, the signal collection scenarios in each environment were categorized into two types, namely static and dynamic.

\textbf{Static scenario}: In each experimental environment, the Wi-Fi signal transmitters were fixed at locations L1, L2, and L3. Specifically, Locations L1 and L2 satisfy the LOS channel condition, while L3 fulfills the NLOS channel condition. 

\textbf{Dynamic scenario}: While the receiver was still placed in the location Rx, the Wi-Fi devices moved randomly at an average speed of about 1 m/s, which is denoted as M in Figs.~\ref{fig:environment_settings}(a)-(c). 

\begin{figure}[!t]
	\centering
	\subfloat[]{\includegraphics[width=1.75in]{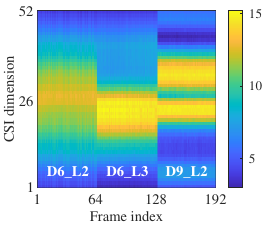}
		\label{fig:CSI_hybrid}}
	\subfloat[]{\includegraphics[width=1.75in]{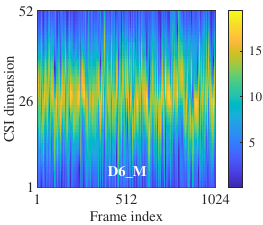}
		\label{fig:CSI_AP6_M}}
	\caption{The estimated CSI from different devices and scenarios in E1 at the same receiving antenna. (a) The estimated CSI from D6 and D9 at L2 and L3. (b) The estimated CSI from D6 in the moving status. }
	\label{fig:CSI}
\end{figure}

The CSI estimated on the active subcarriers from ltf1 in static and mobile scenarios are exemplified in Figs.~\ref{fig:CSI}(a) and \ref{fig:CSI}(b), respectively. As shown in Fig.~\ref{fig:CSI}(a), even in a static scenario, different signals emitted by the same Wi-Fi device located at the same position over a period of time would have a slight channel difference, due to the subtle changes in the environment. 
In the dynamic scenario shown in Fig.~\ref{fig:CSI}(b), since the occasional block of the direct LOS path, and the variations of the device position, the wireless channel conditions between the receiver and transmitters would be significantly changing during the process of data collection.

\subsubsection{Datasets}
In the experiments, a total of 12 datasets (4 per environment) were collected. The primary purpose is to test the effectiveness of the proposed channel-robust RFFI mechanism against drastic channel variations. 

Moreover, in order to verify the effectiveness of the proposed identification method against CFO drift, the datasets from E1 were collected in Feb. 2024 (winter for the northern hemisphere), while the datasets from E2 and E3 were collected in Jul. 2024 (summer), where the temperature-relevant CFO would be different. Besides, the five-month time span also poses a challenge to the consistency and validity of the extracted RFF representations.


An intermittent sampling process was adopted. Specifically, each sampling time was set to 5 seconds, followed by a period of time for data saving. Moreover, there were $10\times512\times4$ samples from L1, L2, and L3, which means that the 4-channel receiver collected 512 frames of the transmitted signal from 10 Wi-Fi devices, respectively, while in the dynamic scenario, the number increased to $10\times1024\times4$. The signal-to-noise ratio (SNR) of the received signals was over 28~dB.

\subsubsection{Training of Neural Networks}
For the CNN-based classification model mentioned in Section~\ref{sec:collaborative classification}, we selected Adam as the optimizer, where the initial learning rate and the weight decay were set to $10^{-3}$ and $10^{-4}$, respectively. The learning rate was reduced by a factor of 0.5 at the end of each epoch, and the batch size was set to 64.

\subsubsection{Algorithm Configurations}
In each evaluation, a dataset collected at a dynamic scenario (i.e., location M) from an environment was selected as the training set, while the datasets acquired from the other two environments were chosen as the test set.
During the training stage, all the signal samples from the 4-antenna receiver were fed into the CNN-based classifier.

In the inference stage, we carried out the ablation studies to evaluate the effect of the following methods.
\begin{itemize}
    \item \textbf{Hybrid}: The proposed CFO-assisted collaborative identification method in this paper;
    \item \textbf{DF (Decision Fusion)}: collaborative identification method without considering CFO;
    \item \textbf{Direct}: identification via single-antenna signal (without collaborative approach) and without considering CFO.
\end{itemize}

\subsection{Evaluation of Proposed Method}\label{performancevschannel}
\subsubsection{Classification Accuracy of the Proposed Method}
The classification results of the proposed RFFI method can be found in the leftmost column in Table~\ref{tab:hybrid_classification_results}, where an average accuracy of 92.76\% is achieved among all the 24 cases.
Notably, when the classification model is trained and tested with datasets from dynamic scenarios in different environments (e.g., trained with E1-M and tested with E2-M), the average classification accuracy reaches 93.38\%, which means that the proposed RFFI method remains consistently effective under the most drastic channel variations.
\begin{table*}[!t]
    \centering
    \caption{Experimental Results}
    \begin{tabular}{|c|c|c|c|c|c|c|c|c|c|c|c|c|c|}
        \hline
        \multirow{3}[0]{*}{Train} & \multirow{3}[0]{*}{Test} & \multicolumn{12}{|c|}{Classification Accuracy (\%)} \\
        \cline{3-14}
        &       & \multicolumn{3}{|c|}{SR} & \multicolumn{3}{|c|}{AS} & \multicolumn{3}{|c|}{DoLoS} & \multicolumn{3}{|c|}{EQ} \\
        \cline{3-14}
        &       & \multicolumn{1}{c}{Hybrid} & \multicolumn{1}{|c|}{DF} & \multicolumn{1}{|c|}{Direct} & \multicolumn{1}{|c|}{Hybrid} & \multicolumn{1}{|c|}{DF} & \multicolumn{1}{|c|}{Direct} & \multicolumn{1}{|c|}{Hybrid} & \multicolumn{1}{|c|}{DF} & \multicolumn{1}{|c|}{Direct} & \multicolumn{1}{|c|}{Hybrid} & \multicolumn{1}{|c|}{DF} & \multicolumn{1}{|c|}{Direct} \\
        \hline
        \multirow{8}[0]{*}{E1-M} & \multicolumn{1}{|c|}{E2-L1} & \textbf{92.05} & \textbf{87.66} & \textbf{76.40} & 84.63 & 73.65 & 56.17 & 73.89 & 68.20 & 50.88 & 76.62 & 62.15 & 50.53 \\
        & \multicolumn{1}{|c|}{E2-L2} & \textbf{95.12} & \textbf{83.79} & \textbf{70.37} & 90.61 & 75.29 & 59.41 & 74.49 & 63.85 & 47.72 & 81.84 & 54.98 & 46.44 \\
        & \multicolumn{1}{|c|}{E2-L3} & 83.83 & 60.29 & 50.49 & \textbf{91.91} & \textbf{82.42} & \textbf{64.33} & 76.50 & 72.13 & 47.29 & 73.05 & 50.60 & 42.75 \\
        & \multicolumn{1}{|c|}{E2-M} & \textbf{86.51} & \textbf{71.46} & \textbf{61.60} & 84.57 & 69.85 & 57.81 & 81.62 & 66.25 & 50.77 & 79.70 & 52.14 & 44.15 \\
        \cline{2-14}
        & \multicolumn{1}{|c|}{E3-L1} & \textbf{88.77} & \textbf{80.92} & \textbf{68.85} & 63.48 & 46.97 & 41.17 & 65.94 & 65.45 & 51.15 & 77.77 & 76.60 & 60.48 \\
        & \multicolumn{1}{|c|}{E3-L2} & \textbf{89.65} & \textbf{81.54} & \textbf{67.01} & 55.10 & 46.23 & 35.78 & 74.02 & 63.40 & 48.08 & 82.19 & 74.14 & 59.20 \\
        & \multicolumn{1}{|c|}{E3-L3} & 89.67 & \textbf{81.86} & \textbf{65.98} & 64.14 & 41.15 & 33.07 & 78.57 & 62.09 & 48.45 & \textbf{89.92} & 75.35 & 60.21 \\
        & \multicolumn{1}{|c|}{E3-M} & 90.75 & \textbf{80.89} & \textbf{70.63} & 71.01 & 51.93 & 43.36 & 79.73 & 63.41 & 51.57 & \textbf{91.04} & 75.13 & 61.30 \\
        \hline
        \multirow{8}[0]{*}{E2-M} & \multicolumn{1}{|c|}{E1-L1} & \textbf{98.12} & \textbf{94.73} & \textbf{78.10} & 96.13 & 90.51 & 72.71 & 74.10 & 68.67 & 48.00 & 81.70 & 64.92 & 56.49 \\
        & \multicolumn{1}{|c|}{E1-L2} & \textbf{98.40} & \textbf{85.14} & \textbf{72.69} & 89.38 & 75.57 & 68.54 & 86.16 & 71.43 & 53.45 & 89.63 & 65.74 & 57.48 \\
        & \multicolumn{1}{|c|}{E1-L3} & 98.09 & 85.29 & 68.21 & \textbf{99.36} & \textbf{88.54} & \textbf{69.61} & 92.03 & 72.52 & 52.05 & 82.68 & 68.48 & 55.63 \\
        & \multicolumn{1}{|c|}{E1-M} & 97.27 & 88.02 & 73.96 & \textbf{97.36} & \textbf{88.38} & \textbf{75.20} & 88.13 & 71.25 & 55.13 & 83.06 & 64.82 & 57.84 \\
        \cline{2-14}
        & \multicolumn{1}{|c|}{E3-L1} & \textbf{95.23} & \textbf{81.02} & \textbf{67.21} & 85.25 & 61.91 & 47.45 & 80.29 & 68.46 & 51.59 & 89.02 & 63.83 & 54.06 \\
        & \multicolumn{1}{|c|}{E3-L2} & \textbf{95.04} & \textbf{88.16} & \textbf{74.35} & 90.33 & 54.98 & 44.52 & 85.25 & 66.56 & 50.39 & 89.18 & 69.82 & 58.12 \\
        & \multicolumn{1}{|c|}{E3-L3} & \textbf{96.70} & \textbf{86.60} & \textbf{70.64} & 92.73 & 62.44 & 50.46 & 85.51 & 65.59 & 49.89 & 95.08 & 73.63 & 58.77 \\
        & \multicolumn{1}{|c|}{E3-M} & 95.17 & \textbf{86.89} & \textbf{75.71} & 87.26 & 61.35 & 50.84 & 91.44 & 68.99 & 55.64 & \textbf{95.53} & 74.78 & 62.51 \\
        \hline
        \multirow{8}[0]{*}{E3-M} & \multicolumn{1}{|c|}{E1-L1} & \textbf{92.30} & \textbf{83.77} & \textbf{71.62} & 77.03 & 67.03 & 55.41 & 68.52 & 67.62 & 51.85 & 72.95 & 79.28 & 62.47 \\
        & \multicolumn{1}{|c|}{E1-L2} & \textbf{96.62} & \textbf{87.50} & \textbf{71.35} & 71.21 & 50.31 & 42.56 & 82.70 & 78.54 & 59.36 & 84.73 & 77.99 & 59.61 \\
        & \multicolumn{1}{|c|}{E1-L3} & \textbf{93.57} & \textbf{82.75} & \textbf{64.58} & 78.55 & 65.27 & 49.77 & 81.31 & 68.67 & 49.14 & 80.63 & 70.66 & 53.37 \\
        & \multicolumn{1}{|c|}{E1-M} & \textbf{95.84} & \textbf{86.53} & \textbf{71.06} & 81.92 & 66.98 & 54.46 & 84.44 & 71.21 & 54.36 & 85.54 & 73.82 & 58.73 \\
        \cline{2-14}
        & \multicolumn{1}{|c|}{E2-L1} & \textbf{92.11} & \textbf{84.34} & \textbf{72.49} & 85.18 & 58.34 & 47.28 & 78.34 & 67.25 & 50.02 & 78.11 & 67.64 & 56.93 \\
        & \multicolumn{1}{|c|}{E2-L2} & \textbf{81.95} & \textbf{76.23} & \textbf{63.87} & 73.77 & 56.11 & 42.87 & 69.98 & 64.75 & 46.78 & 72.68 & 63.42 & 53.55 \\
        & \multicolumn{1}{|c|}{E2-L3} & \textbf{88.73} & \textbf{71.02} & \textbf{53.21} & 86.99 & 65.18 & 50.46 & 80.90 & 66.62 & 45.27 & 88.20 & 61.60 & 46.82 \\
        & \multicolumn{1}{|c|}{E2-M} & \textbf{94.75} & \textbf{79.34} & \textbf{66.35} & 88.06 & 62.04 & 49.75 & 88.69 & 72.48 & 55.11 & 90.92 & 65.63 & 54.43 \\
        \hline
        \multicolumn{2}{|c|}{Average} & \textbf{92.76} & \textbf{82.32} & \textbf{68.61} & 82.75 & 65.10 & 52.62 & 80.11 & 68.14 & 51.00 & 83.82 & 67.80 & 55.49 \\
        \hline
    \end{tabular}%
    \label{tab:hybrid_classification_results}%
\end{table*}%

\subsubsection{Contribution of Decision Fusion}
As presented in Table~\ref{tab:hybrid_classification_results}, the DF method exhibits great effectiveness and generalizability. Compared with the Direct methods, the proposed DF method achieves an average accuracy improvement of 12.30\% to 17.14\% for recognition results based on different RFF representations. Due to the fact that multi-antenna AP routers are common in practical Wi-Fi communication scenarios, the DF algorithm has significant value in real-world applications.    

\subsection{Evaluation of RFF Representations}    
\subsubsection{Comparison Between Spectral Regrowth (SR) and Spectrum on Active Subcarriers (AS)}
As shown in Table~\ref{tab:hybrid_classification_results}, regardless of the employed classification algorithm, the average accuracy obtained by the SR-based method exceeds that of the AS-based method by more than 10.01\% across the 24 cases, which demonstrates a greater potential against channel variations for SR-based RFF representation.

When the training and test sets are sourced from office environments E1 and E2, it is observed that the AS-based method also achieves satisfactory results. However, when the environmental change is more drastic, for example, when E1-M is used as the training set and the dataset from E3 is used as the test set, the classification outcomes deteriorate significantly. Consequently, it can be inferred that the features learned by the classifier utilizing the AS-based RFF representation remain selective to the channel environment, prone to overfitting in cases of drastic environmental variations, and therefore exhibit far inferior generalizability compared to the SR-based RFF representation.

Furthermore, from Section~\ref{sec:simulation}, it can be inferred that the RFF representation on active subcarriers has the following disadvantages compared to the spectral regrowth:

\begin{itemize}
	\item
	It is difficult for the network to learn the nonlinear features due to the overwhelming presence of memory effects (mostly $A_{1}(\omega)$) on the active subcarriers.
\end{itemize}
\begin{itemize}
	\item
	From (\ref{eq:Yiomega}) and (\ref{eq:Somega}), both the channel impacts $H(\omega)$ and memory effect $A_{1}(\omega)$ are multiplicative to the spectrum, thus the features learned by the network are more prone to residual channel interference, leading to overfitting issues.
\end{itemize}
\begin{itemize}
	\item
	Due to the existence of $A_{1}(\omega)U(\omega)$, the absolute value of amplitude variations induced by channel impacts on active subcarriers would be significantly larger than those on inactive ones, which may further exacerbate the difficulty for the network to learn channel-resistant features.
\end{itemize}

Therefore, the spectral regrowth is more suitable as an RFF representation to be fed into the neural network classifier for channel-resistant RFFI.

\subsubsection{Comparison Between Spectral Regrowth and Other RFF Representations}
The following two RFF representations from the literature are chosen for benchmarking. 
\begin{itemize}
    \item DoLoS (difference of the logarithm of the spectrum)~\cite{9979789}: the logarithmic spectrums of two adjacent symbols were used to extract channel-robust RFF. 
    \item EQ (signal after channel equalization)~\cite{8882379}: the channel equalization was used to eliminate the channel impacts.     
\end{itemize}

For fairness, the preamble is utilized to extract the RFF, and the Hybrid method is employed for classification in each case. All results are averaged over multiple experiments.

As demonstrated in Table~\ref{tab:hybrid_classification_results}, it is evident that the SR-based identification method proposed in this paper is far superior to that based on DoLoS or EQ in most cases, with an average accuracy improvement of 12.65\% and 8.94\%, respectively.

It is also worth mentioning that when the classification scheme turns into DF and Direct as shown in Table~\ref{tab:hybrid_classification_results}, the SR-based recognition results achieve the highest accuracy in 21 out of 24 cases, with an average accuracy improvement of at least 13.12\% compared to the others.

\subsection{Evaluation of the CFO Assisted Mechanism}\label{sec:performance_CFO}

\subsubsection{Visualization of CFO Drift}

\begin{figure}[!t]
    \centering
    \subfloat[]{\includegraphics[width=1.75in]{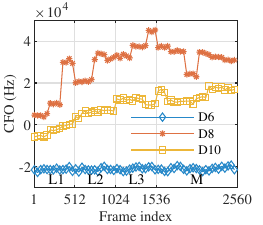}
        \label{fig:CFO_2device_timevaring}}
    \subfloat[]{\includegraphics[width=1.75in]{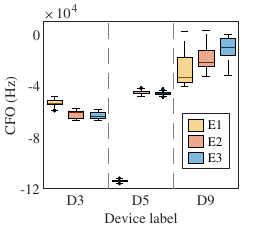}}
    \caption{CFO drift in datasets. (a) CFO drift of D6, D8, and D10 at E1-L1, L2, L3, and M. (b) CFO distribution of D3, D5, and D9 in E1, E2, and E3.}
    \label{fig:CFOdrift}
\end{figure}

Due to the oscillator's sensitivity to temperature variations, the estimated CFO may exhibit time-varying characteristics. As illustrated in Fig.~\ref{fig:CFOdrift}(a), significant time-varying characteristics and fluctuations manifested as irregular jumps can be observed on the estimated CFO of D8 and D10. This phenomenon is mainly caused by the temperature variation during the device operation, together with the intermittent acquisition of the signal. However, for some devices, e.g., D6, the estimated CFO values remain relatively stable over time, indicating that the degree of CFO drift varies among different devices.

Furthermore, on a longer time scale, as shown in Fig.~\ref{fig:CFOdrift}(b), although the range of estimated CFO in the E1 dataset is quite different in comparison to that in E2 and E3, the CFO ranges in the E2 and E3 datasets are relatively similar. Moreover, the CFO variation is limited within a given environment. This suggests that the CFO drift is limited over several days, since the E2 and E3 datasets were collected in the same month while E1 dataset was collected five months apart.

Besides, from Figs.~\ref{fig:CFOdrift}(a) and (b), although the degree of CFO drift varies from device to device, the estimated CFO values are still roughly distinguishable among devices. Therefore, CFOs can be utilized to aid the identification process.

\subsubsection{Contribution of CFO}
As presented in Table~\ref{tab:hybrid_classification_results}, an average improvement of 10.44\% for the SR-based identification result can be found when the CFO weights are employed in the DF mechanism. Therefore, the estimated CFO values do have a positive effect on the classification results.
\begin{figure}[!t]
	\centering
	\subfloat[]{\includegraphics[width=1.16in]{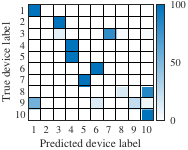}}
	\subfloat[]{\includegraphics[width=1.16in]{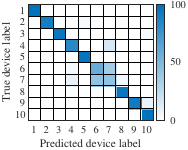}}
	\subfloat[]{\includegraphics[width=1.16in]{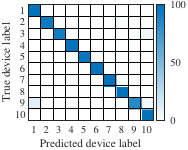}}\\
	\subfloat[]{\includegraphics[width=1.16in]{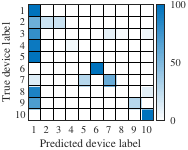}}
	\subfloat[]{\includegraphics[width=1.16in]{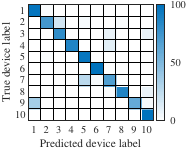}}
	\subfloat[]{\includegraphics[width=1.16in]{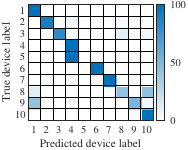}}
	\caption{Classification results when trained with E2-M and tested with E1-M achieved by (a) only CFO weights: 45.05\%, (b) SR-based RFF, DF: 88.02\%, (c) SR-based RFF, Hybrid: 97.27\%, (d) SR-based RFF, DF, and \cite{9448147}: 41.53\%, (e) SR-based RFF, DF, and the improved \cite{9448147}: 85.02\%, (f) SR-based RFF, DF, and UD: 76.85\%.}
	\label{fig:CFO CM}
\end{figure}

The proposed CFO-assisted algorithm also presents outstanding resistance to CFO drift. When E2-M is the training set and E1-M is the test set, as shown in Fig.~\ref{fig:CFO CM}(a), a classification accuracy of 45.05\% is obtained with CFO weights alone. 
This phenomenon is mainly attributed to the temperature difference between the two datasets, where E1 is collected in winter while E2 is in summer. However, when the CFO weights are employed in the DF method as described in Section~\ref{sec:decision fusion}, a final accuracy of 97.27\% is obtained as shown in Fig.~\ref{fig:CFO CM}(c), with an improvement of 9.25\% compared to the result of DF method. Interestingly, from Figs.~\ref{fig:CFO CM}(a)-(c), it can be seen that although some devices (e.g., D7), either with CFO weights alone or with DF method, are not quite accurate, a final Hybrid classification result close to 100\% is still obtained. This is because the CFO, while it may have significant drift, the range of drift is still generally limited, thus the CFO value can still provide an auxiliary recognition effect.

Specifically, for D5, it can be seen from Fig.~\ref{fig:CFOdrift}(b) that its CFO drift is severe, however, due to the strong distinguishability of its SR-based RFF representation, the CFO weight does not affect its Hybrid classification result. For devices with slightly lower differentiability, such as D9, the effect of CFO drift is also reflected in the final recognition results, where a small number of samples from D9 to be tested are classified as D1 as shown in Fig.~\ref{fig:CFO CM}(c).

\subsubsection{Comparison with Related CFO Assisted Methods}
We also compared our approach with other CFO-assisted methods in the literature.
\begin{itemize}
    \item In \cite{9448147}, the estimated CFO was utilized to set the prediction probability to zero when it was much different from the reference one. 
    \item We designed an improved algorithm based on \cite{9448147}. When all the prediction probabilities are set to zero, the prediction of this tested sample will be done by the SR-based RFF alone. The CFO drift of Wi-Fi devices is much larger than that of LoRa studied in \cite{9448147}, thus a default prediction should be added to the algorithm.
    \item Moreover, an undercomplete demodulation (UD) algorithm was proposed in \cite{8882379}, where the estimated CFO was reintroduced to the signal after preprocessing.

\end{itemize}

After replacing the CFO-assisted algorithm in this paper with the above three methods, the classification results are shown in Figs.~\ref{fig:CFO CM}(d)-(f). It is evident that all three methods show a disadvantage in the face of CFO drift, and none of them exceeds the accuracy obtained without CFO assistance which is shown in Fig.~\ref{fig:CFO CM}(c).

\section{Conclusion}\label{sec:conclusion}
In this paper, we proposed a channel-robust RFFI system based on spectral regrowth and CFO, and carried out extensive experimental evaluations. In detail, the spectral regrowth was utilized as an RFF representation which is rooted in the PA nonlinearity. Moreover, the CFO was used as an auxiliary feature to calibrate the deep learning-based inference. Although CFO exhibits time-varying characteristics, the range of CFO drift was found to be limited over several days, and a rough distinguishability among devices was also observed over several months. Finally, a CNN-based collaborative classification method was employed to utilize the diversity of the multi-antenna receiver.
We performed experiments under various practical communication environments and a collection time span of 5 months, involving 10 commercial Wi-Fi devices and a USRP X310 receiver with 4 antennas. Moreover, static scenarios under LOS and NLOS configurations, and dynamic scenarios with moving transmitters were considered. The results demonstrated the effectiveness of the spectral regrowth-based RFF representation, where an average accuracy of 92.76\% was achieved when faced with channel variations, far superior to the related methods with an improvement of at least 8.94\%. Furthermore, the results demonstrated the robustness of the proposed CFO-assisted collaborative classification model against CFO drift, where the recognition accuracy was improved by 9.25\% under a 5-month time span, while the other benchmark methods failed to obtain positive effects.



\bibliographystyle{IEEEtran}

\begin{thebibliography}{10}
\providecommand{\url}[1]{#1}
\csname url@samestyle\endcsname
\providecommand{\newblock}{\relax}
\providecommand{\bibinfo}[2]{#2}
\providecommand{\BIBentrySTDinterwordspacing}{\spaceskip=0pt\relax}
\providecommand{\BIBentryALTinterwordstretchfactor}{4}
\providecommand{\BIBentryALTinterwordspacing}{\spaceskip=\fontdimen2\font plus
\BIBentryALTinterwordstretchfactor\fontdimen3\font minus \fontdimen4\font\relax}
\providecommand{\BIBforeignlanguage}[2]{{%
\expandafter\ifx\csname l@#1\endcsname\relax
\typeout{** WARNING: IEEEtran.bst: No hyphenation pattern has been}%
\typeout{** loaded for the language `#1'. Using the pattern for}%
\typeout{** the default language instead.}%
\else
\language=\csname l@#1\endcsname
\fi
#2}}
\providecommand{\BIBdecl}{\relax}
\BIBdecl

\bibitem{9961104}
A.~I. Siam, M.~A. El-Affendi, A.~A. Elazm, G.~M. El-Banby, N.~A. El-Bahnasawy, F.~E.~A. El-Samie, and A.~A.~A. El-Latif, ``Portable and real-time {IoT}-based healthcare monitoring system for daily medical applications,'' \emph{IEEE Trans. Comput. Soc. Syst.}, vol.~10, no.~4, pp. 1629--1641, 2023.

\bibitem{9195014}
M.~Serror, S.~Hack, M.~Henze, M.~Schuba, and K.~Wehrle, ``Challenges and opportunities in securing the industrial internet of things,'' \emph{IEEE Trans. Ind. Inform.}, vol.~17, no.~5, pp. 2985--2996, 2021.

\bibitem{9149584}
J.~Zhang, G.~Li, A.~Marshall, A.~Hu, and L.~Hanzo, ``A new frontier for {IoT} security emerging from three decades of key generation relying on wireless channels,'' \emph{IEEE Access}, vol.~8, pp. 138\,406--138\,446, 2020.

\bibitem{8737455}
P.~Liu, P.~Yang, W.~Song, Y.~Yan, and X.~Li, ``Real-time identification of rogue {WiFi} connections using environment-independent physical features,'' in \emph{Proc. IEEE INFOCOM}, 2019, pp. 190--198.

\bibitem{refId0}
L.~Xie, L.~Peng, J.~Zhang, and A.~Hu, ``Radio frequency fingerprint identification for internet of things: A survey,'' \emph{Secur. Saf.}, vol.~3, p. 2023022, 2024.

\bibitem{9517121}
R.~Xie, W.~Xu, Y.~Chen, J.~Yu, A.~Hu, D.~W.~K. Ng, and A.~L. Swindlehurst, ``A generalizable model-and-data driven approach for open-set {RFF} authentication,'' \emph{IEEE Trans. Inf. Forensic Secur.}, vol.~16, pp. 4435--4450, 2021.

\bibitem{10508246}
P.~Tang, G.~Ding, Y.~Xu, Y.~Jiao, Y.~Song, and G.~Wei, ``Causal learning for robust specific emitter identification over unknown channel statistics,'' \emph{IEEE Trans. Inf. Forensic Secur.}, vol.~19, pp. 5316--5329, 2024.

\bibitem{7464336}
T.~J. Bihl, K.~W. Bauer, and M.~A. Temple, ``Feature selection for {RF} fingerprinting with multiple discriminant analysis and using {ZigBee} device emissions,'' \emph{IEEE Trans. Inf. Forensic Secur.}, vol.~11, no.~8, pp. 1862--1874, 2016.

\bibitem{9715147}
G.~Shen, J.~Zhang, A.~Marshall, and J.~R. Cavallaro, ``Towards scalable and channel-robust radio frequency fingerprint identification for {LoRa},'' \emph{IEEE Trans. Inf. Forensic Secur.}, vol.~17, pp. 774--787, 2022.

\bibitem{9965952}
M.~Nair, T.~A. Cappello, S.~Dang, and M.~A. Beach, ``Rigorous analysis of data orthogonalization for self-organizing maps in machine learning cyber intrusion detection for {LoRa} sensors,'' \emph{IEEE Trans. Microw. Theory Tech.}, vol.~71, no.~1, pp. 389--408, 2023.

\bibitem{10050441}
J.~He, S.~Huang, S.~Chang, F.~Wang, B.~Shen, and Z.~Feng, ``Radio frequency fingerprint identification with hybrid time-varying distortions,'' \emph{IEEE Trans. Wirel. Commun.}, vol.~22, no.~10, pp. 6724--6736, 2023.

\bibitem{10430094}
H.~Fu, Y.~Sun, L.~Peng, and M.~Liu, ``Channel-resilient {RF} fingerprint identification based on nonlinear features with memory effect,'' \emph{IEEE Commun. Lett.}, vol.~28, no.~4, pp. 798--802, 2024.

\bibitem{9979789}
Y.~Xing, A.~Hu, J.~Zhang, L.~Peng, and X.~Wang, ``Design of a channel robust radio frequency fingerprint identification scheme,'' \emph{IEEE Internet Things J.}, vol.~10, no.~8, pp. 6946--6959, 2023.

\bibitem{9335608}
M.~Fadul, D.~Reising, T.~D. Loveless, and A.~Ofoli, ``Nelder-mead simplex channel estimation for the {RF-DNA} fingerprinting of {OFDM} transmitters under {Rayleigh} fading conditions,'' \emph{IEEE Trans. Inf. Forensic Secur.}, vol.~16, pp. 2381--2396, 2021.

\bibitem{8882379}
K.~Sankhe, M.~Belgiovine, F.~Zhou, L.~Angioloni, F.~Restuccia, S.~D’Oro, T.~Melodia, S.~Ioannidis, and K.~Chowdhury, ``No radio left behind: Radio fingerprinting through deep learning of physical-layer hardware impairments,'' \emph{IEEE Trans. Cogn. Commun. Netw.}, vol.~6, no.~1, pp. 165--178, 2020.

\bibitem{9155259}
A.~Al-Shawabka, F.~Restuccia, S.~D’Oro, T.~Jian, B.~Costa~Rendon, N.~Soltani, J.~Dy, S.~Ioannidis, K.~Chowdhury, and T.~Melodia, ``Exposing the fingerprint: Dissecting the impact of the wireless channel on radio fingerprinting,'' in \emph{Proc. IEEE INFOCOM}, Toronto, ON, Canada, July 2020, pp. 646--655.

\bibitem{10332157}
Y.~Shao, J.~Liu, Y.~Zeng, and Y.~Gong, ``A radio frequency fingerprinting scheme using learnable signal representation,'' \emph{IEEE Commun. Lett.}, vol.~28, no.~1, pp. 73--77, 2024.

\bibitem{10360105}
X.~Yang and D.~Li, ``{LED-RFF}: {LTE} {DMRS}-based channel robust radio frequency fingerprint identification scheme,'' \emph{IEEE Trans. Inf. Forensic Secur.}, vol.~19, pp. 1855--1869, 2024.

\bibitem{10511278}
Y.~Peng, X.~Zhang, L.~Guo, C.~Ben, Y.~Liu, Y.~Wang, Y.~Lin, and G.~Gui, ``Enhanced specific emitter identification with limited data through dual implicit regularization,'' \emph{IEEE Internet Things J.}, early access, Apr. 30, 2024, doi: 10.1109/JIOT.2024.3395441.

\bibitem{10285131}
C.~Liu, X.~Fu, Y.~Wang, L.~Guo, Y.~Liu, Y.~Lin, H.~Zhao, and G.~Gui, ``Overcoming data limitations: A few-shot specific emitter identification method using self-supervised learning and adversarial augmentation,'' \emph{IEEE Trans. Inf. Forensic Secur.}, vol.~19, pp. 500--513, 2024.

\bibitem{10107741}
A.~Jagannath and J.~Jagannath, ``Embedding-assisted attentional deep learning for real-world {RF} fingerprinting of {Bluetooth},'' \emph{IEEE Trans. Cogn. Commun. Netw.}, vol.~9, no.~4, pp. 940--949, 2023.

\bibitem{10229972}
Q.~Zhou, Y.~He, K.~Yang, and T.~Chi, ``Physical-layer identification of wireless {IoT} nodes through {PUF}-controlled transmitter spectral regrowth,'' \emph{IEEE Trans. Microw. Theory Tech.}, vol.~72, no.~2, pp. 1045--1055, 2024.

\bibitem{9348261}
G.~Reus-Muns, D.~Jaisinghani, K.~Sankhe, and K.~R. Chowdhury, ``Trust in {5G} open {RANs} through machine learning: {RF} fingerprinting on the {POWDER PAWR} platform,'' in \emph{Proc. IEEE Glob. Commun. Conf.}, Taipei, Taiwan, Dec. 2020, pp. 1--6.

\bibitem{5740592}
M.~Liu and J.~F. Doherty, ``Nonlinearity estimation for specific emitter identification in multipath channels,'' \emph{IEEE Trans. Inf. Forensic Secur.}, vol.~6, no.~3, pp. 1076--1085, 2011.

\bibitem{1703853}
D.~Morgan, Z.~Ma, J.~Kim, M.~Zierdt, and J.~Pastalan, ``A generalized memory polynomial model for digital predistortion of {RF} power amplifiers,'' \emph{IEEE Trans. Signal Process.}, vol.~54, no.~10, pp. 3852--3860, 2006.

\bibitem{7122344}
A.~C. Polak and D.~L. Goeckel, ``Identification of wireless devices of users who actively fake their {RF} fingerprints with artificial data distortion,'' \emph{IEEE Trans. Wirel. Commun.}, vol.~14, no.~11, pp. 5889--5899, 2015.

\bibitem{6804410}
W.~Hou, X.~Wang, J.-Y. Chouinard, and A.~Refaey, ``Physical layer authentication for mobile systems with time-varying carrier frequency offsets,'' \emph{IEEE Trans. Commun.}, vol.~62, no.~5, pp. 1658--1667, 2014.

\bibitem{8485917}
J.~Hua, H.~Sun, Z.~Shen, Z.~Qian, and S.~Zhong, ``Accurate and efficient wireless device fingerprinting using channel state information,'' in \emph{Proc. IEEE INFOCOM}, Honolulu, HI, USA, Apr. 2018, pp. 1700--1708.

\bibitem{10184130}
H.~Fu, L.~Peng, M.~Liu, and A.~Hu, ``Deep learning-based {RF} fingerprint identification with channel effects mitigation,'' \emph{IEEE Open J. Commun. Soc.}, vol.~4, pp. 1668--1681, 2023.

\bibitem{9448147}
G.~Shen, J.~Zhang, A.~Marshall, L.~Peng, and X.~Wang, ``Radio frequency fingerprint identification for {LoRa} using deep learning,'' \emph{IEEE J. Sel. Areas Commun.}, vol.~39, no.~8, pp. 2604--2616, 2021.

\bibitem{9450821}
J.~Zhang, R.~Woods, M.~Sandell, M.~Valkama, A.~Marshall, and J.~Cavallaro, ``Radio frequency fingerprint identification for narrowband systems, modelling and classification,'' \emph{IEEE Trans. Inf. Forensic Secur.}, vol.~16, pp. 3974--3987, 2021.

\bibitem{8031977}
B.~Bloessl, M.~Segata, C.~Sommer, and F.~Dressler, ``Performance assessment of {IEEE} 802.11p with an open source {SDR}-based prototype,'' \emph{IEEE. Trans. Mob. Comput.}, vol.~17, no.~5, pp. 1162--1175, 2018.

\bibitem{8798719}
Y.~Chen, X.~Su, Y.~Hu, and B.~Zeng, ``Residual carrier frequency offset estimation and compensation for commodity {WiFi},'' \emph{IEEE. Trans. Mob. Comput.}, vol.~19, no.~12, pp. 2891--2902, 2020.

\bibitem{651189}
G.~Raz and B.~van Veen, ``Baseband {Volterra} filters for implementing carrier based nonlinearities,'' \emph{IEEE Trans. Signal Process.}, vol.~46, no.~1, pp. 103--114, 1998.

\end{thebibliography}

\end{document}